\documentclass{ws-procs9x6}
\usepackage{epstopdf}

\def\etal {{\it et al.}}

\begin{document}

\title{USING A ROTATING MAGNETIC GUIDING FIELD FOR THE $^3$He-$^{129}$Xe-COMAGNETOMETER}

\author{F. \ ALLMENDINGER$^*$ and U.\ SCHMIDT}

\address{Physikalisches Institut, Ruprecht-Karls-Universit\"at Heidelberg,\\
69120 Heidelberg, Germany\\
$^*$E-mail: allmendinger@physi.uni-heidelberg.de}

\author{W.\ HEIL, S.\ KARPUK, A.\ SCHARTH, Y.\ SOBOLEV,  K.\ TULLNEY and S.\ ZIMMER}

\address{Institut f\"{u}r Physik, Johannes Gutenberg-Universit\"{a}t,\\ 
55099 Mainz, Germany}

\begin{abstract}
Our search for non-magnetic spin-dependent interactions is based on the measurement of free precession of nuclear spin polarized $^3$He and $^{129}$Xe atoms in a homogeneous magnetic guiding field of about 400 nT. We report on our approach to perform an adiabatic rotation of the guiding field that allows us to modulate possible non-magnetic spin-dependent interactions and to find an optimization procedure for long transverse relaxation times $T_2^*$ both for Helium and Xenon.
\end{abstract}

\bodymatter
\section{Introduction}
The most precise tests of non-magnetic spin-dependent interactions like (i) the Lorentz invariance violating coupling of spins to a background field\cite{datatables} or (ii) the search for a P- and T-violating short-range interaction mediated by light, pseudo-scalar bosons (Axions)\cite{Tullney} are often performed in experiments that compare the transition frequencies of two co-located systems (clock comparison)\cite{Schmidt,Gemmel, Gemmel2}. This type of experiment is only sensitive if the spin-dependent interaction in question is varying in time. In case of the search for a coupling of spins to a relic background field \cite{Schmidt , Gemmel}, the Earth's daily rotation leads to a modulation of the orientation in space with the sidereal frequency. As the frequency of this modulation is unfavorably low, we developed a system to rotate the magnetic guiding field of our $^3$He-$^{129}$Xe-comagnetometer ($\Omega_{rot}\approx 2\pi / (20 \text{ min})$) in order to have a much more preferable modulation frequency that may help to get rid of the correlated errors which at present limit our overall sensitivity. Inside the $\mu$-metal shielded room BMSR-2 at PTB, Berlin, two square Helmholtz coil pairs (fixed perpendicular to each other) are driven by low noise, high resolution current sources in order to provide the homogeneous magnetic field of about 400 nT. The stability of the current is $\Delta I/I\approx 10^{-6}$, while the stability of the magnetic field is $\Delta B/B\approx 10^{-4}$ due to drifts of the $\mu$-metal shielding, measured by analyzing Larmor frequency drifts. With this setup we can generate a guiding field, which points into any direction $\alpha$ in the horizontal plane.
\begin{figure}
\begin{center}
\psfig{file=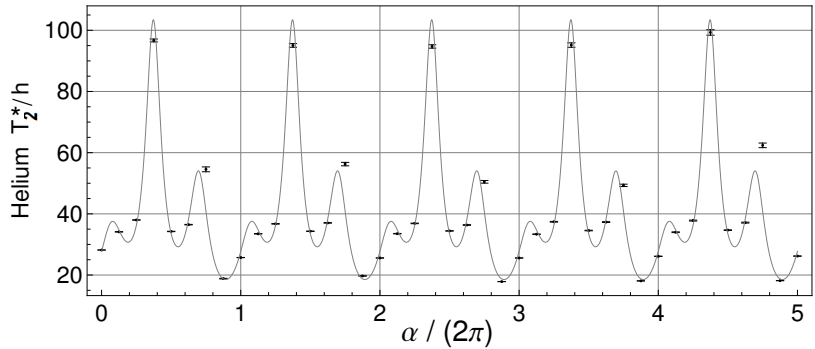,width=4.4in}
\end{center}
\caption{The transverse relaxation time of Helium as a function of the direction $\alpha$ of the magnetic guiding field in the horizontal plane, measured in steps of $45^\circ$ for 5 turns. In total, the measurement took  about 20 hours with $\sim$30 min for each field setting to extract the $T_2^*$ from the decay of the signal amplitude. Solid line: Fit of a Fourier series to the relaxation rates. Corresponding Xenon data in Ref. \cite{Heil}.}
\label{aba:fig1}
\end{figure}
\section{Results}
Besides the proof of principle that a coherent measurement of spin precession with a slowly rotating guiding field is possible (i.e., we do not lose the phase information), our results have an unexpected impact on the measurements with static guiding fields: The transverse relaxation time strongly depends on the direction of the magnetic guiding field and varies between 20 h and 100 h for Helium (fig. \ref{aba:fig1}), and between 6 h and 8.5 h for Xenon. The characteristic pattern in fig. \ref{aba:fig1} for $T_{2, He}^*$ repeats itself after every revolution and is reproduced in all successive runs over a period of at least two weeks. For $T_{2, Xe}^*$, the characteristic pattern is similar\cite{Heil}. This effect of "incidental shimming"\cite{Harcken} has the following explanation: The presence of  magnetic field gradients across a sample cell causes an increased transverse relaxation rate. The origin of this relaxation mechanism is the loss of phase coherence. For a spherical sample cell of radius R the relaxation rate $1/T_2^*$ is \cite{Cates}
\begin{equation}
\frac{1}{T_2^*}=\frac{1}{T_1}+\frac{4R^4\gamma ^2}{175D}\left(|\vec{\nabla} B_y|^2+|\vec{\nabla} B_z|^2+2|\vec{\nabla }B_x|^2\right)
\label{aba:eq1}
\end{equation}
with the guiding field pointing into the x-direction. $\gamma$ is the gyromagnetic ratio and D is the diffusion coefficient of the gas. It is useful to measure at low fields in order to minimize the absolute field gradients, which are of order pT/cm inside BMSR-2. There are two main sources of gradients: Residual field gradients from the $\mu$-metal shielding and gradients produced by the Helmholtz coils. The latter ones will change, as the magnetic guiding field is rotated. At some angle $\alpha$  the gradients from the chamber and coils almost cancel each other and $T_2^*$ is maximized. At other angles the cancellation is less distinct with a minimum in $T_2^*$ at a field orientation where the gradients add up constructively. This is consistent with the observation that by rotating the magnetic guiding field by $180^\circ$ the transverse relaxation time goes from the global maximum to the global minimum.
\section{Conclusion}
These results gave us the possibility to optimize $T_2^*$ and thus the observation time for the search for a Lorentz violating coupling of spins to a hypothetical background field. In our present experiments coherent spin precession can be monitored for more than 24 hours ($\approx 3\cdot T_{2, Xe}^*$)\cite{Schmidt}. Longer observation times $T$ cause a higher sensitivity in frequency measurement ($\sigma \propto T^{-3/2}$) according to the Cramer-Rao Lower Bound \cite{Gemmel} and, furthermore, greatly reduce the correlated errors, especially if runs of coherent spin precession are substantially longer than the period of a sidereal day.\cite{Gemmel2}

\end{document}